\documentclass[aps,prl,twocolumn]{revtex4}
\usepackage{graphicx}
\begin{document}

\title{Instabilities of micro-phase separated Coulombic systems in
constant electric fields.}

\author{Francisco J. Solis}

\affiliation{Integrated Natural Sciences, Arizona State University
West, Phoenix, Arizona}
\author{ Galen T. Pickett}
\affiliation{ Physics Department, California State University Long
Beach, Long Beach, California}

\date{\today}

\begin{abstract}
Mixtures of near-symmetric oppositely charged components with
strong attractive short range interactions exhibit ordered
lamellar phases at low temperatures. In the strong segregation
limit the state of these systems can be described by the location
of the interfaces between the components. It has previously been
shown that these systems are stable against small deformations of
the interfaces. We examine their stability in the presence of a
uniform electric field. When the field is perpendicular to the
lamellae, the system is unstable against long wavelength
deformations for all non-zero values of the external field. A
field parallel to the lamellae produces deformed but persistent
interfaces. In a finite thickness system, onset of an external
perpendicular field modifies the ground state. Flow between the
old and new ground states requires the destruction of the original
interfaces; this destruction proceeds through the instabilities
identified in the bulk case. We examine the possibility of
dynamical stabilization of structures by means of oscillating
fields.
\end{abstract}
\maketitle


In binary mixtures, the competition between repulsive short-range,
and attractive long-range interactions leads to micro-phase
segregation and pattern formation. A basic example is a mixture of
immiscible, oppositely charged particles, where electrostatic
interactions frustrate the full phase segregation of the species.
This basic scenario has been considered in contexts such as
charged polymer blends \cite{Galen1}, polyelectrolytes
\cite{mur-polyelec2}, multi-component micelles \cite{solis05},
alloys \cite{mur-ceramics,thorpe2d}, and photostimulated
semiconductors \cite{mur-molten}. These systems can be studied at
a more abstract level\cite{muratov,lorenzana} since concrete
models often turn out to be equivalent to those describing other
systems such as multiblock copolymers
\cite{mur-copoly,mur-diblock} and reaction-diffusion systems
\cite{mur-reaction,mur-reaction2}, among others.

Control or modification of self-assembled structures is of course
of great practical importance. It is natural to consider external
electric fields as
a
mean of morphology control\cite{solis-layers}, as these directly
couple to the local properties of the system and are directly
implementable. Morphology manipulation by electric fields has
already proven successful in block copolymer systems in which the
microphases are not charged but have different dielectric
constants. \cite{surfacePotentials,NanodomainControl,williams} We
investigate in this article the effects of coupling the charged
binary mixture with an external electric field.
We describe the ground state of the system in the presence of the
field; this turns out to also be a lamellar state with suitably
modified thickness at the edges of the system. Our main result is
that the original lamellar geometry, at the onset of a
perpendicular external field, is an unstable equilibrium state of
the system. We use this result and the nature of the perturbation
created by external fields parallel to the lamellae to describe
some of the main features of the dynamics that bridges the
original and final equilibrium state of the system.

The model we discuss operates in the strong segregation limit
where concentration fluctuation effects are negligible, and where
the electrostatic interaction is weak compared to the segregation
tendencies of the mixture.
At each point in space there is a well defined majority component
and charge. Between these oppositely charged regions there are
sharp interfaces and there is a well defined surface tension
$\gamma$ between them.
This approximation is useful when the characteristic thickness of
the interface is much smaller than the characteristic size of the
segregated domains.
The segregated regions are considered large
with respect to the particle size and can be treated as continuous
media with prescribed charge density $\rho_0$.


We can write an effective free energy for the system with
contributions from the interfacial surface tension, the
electrostatic self-energy, and the interaction with the external
field. We consider only the symmetric case where the magnitude of
the charge density $\rho_{0}$ is the same for each of the species.
Dimensional analysis shows that the system has characteristic
length and energy scales given by $l=(\rho_{0}^2/\gamma)^{1/3}$,
and $g=\gamma l^2$. All physical variables are adimentionalized
using these scales. We use a charge indicator field $\rho$ that
only takes the values $\pm 1$. The electric field and potential
generated by the charges in the system are ${\bf E}_i$ and
$\phi_i$, while ${\bf E}_e$ and $\phi_e$ correspond to the
externally imposed field and its associated potential. Using these
conventions, the free energy of the system is
\begin{eqnarray}
F&=&\int_{\Gamma} dS+\frac{1}{2}\int_\Lambda d{\bf x}\int_\Lambda
d{\bf y} \frac{ \rho({\bf x})\rho({\bf y})}{|{\bf x}-{\bf
y}|}\nonumber \\&+&\int_{\Lambda}d{\bf x} \phi_{e}\rho
-\int_{\Lambda}d{\bf x}\,\psi(|\rho|^{-1}-1).
\end{eqnarray}
The region occupied by the charges is labelled $\Lambda$, and the
interfaces within the system are labelled by $\Gamma$. In the last
term we introduce a dimensionless pressure field $\psi$ that acts
as a Lagrangian multiplier enforcing the condition of
incompressibility. This term does not contribute to the net value
of the energy but restricts the possible conformations of the
system and its allowed deformations.

\begin{figure}
\centering
\includegraphics{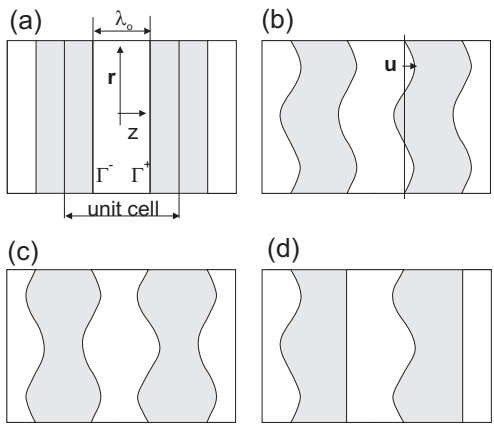}
\caption{\label{fig:lam} Panel (a) shows the ground state lamellar
structure, with layers of width $\lambda_{o}$.
White
regions are positively charged, shaded regions are negative. The
scheme shows our choice of unit cell and coordinate system and the
assignment of labels $\Gamma^{+}$, $\Gamma^{-}$ to the interfaces.
Panels (b-d) show different normal deformation modes; (b) is the
charge antisymmetric or wiggle mode, (c) is the charge symmetric
or corrugation mode, and (d) appears at the edge of the Brillouin
zone.  Panel (b) also shows the surface displacement vector ${\bf
u}$.}
\end{figure}


For components with equal or near equal charge density, the ground
state of the system has a lamellar geometry with alternating
layers of homogenous charge as shown in Fig. \ref{fig:lam}a. In
the symmetric case, in zero external field, the net bulk free
energy per unit volume $f$ for a lamellar pattern of layers of
width $\lambda$ is
\begin{equation}f=\frac{1}{\lambda}+\frac{2 \pi}{3}{\lambda^2}.
\end{equation}
The optimal value of the thickness is then $\lambda_{0}=(3/
\pi)^{1/3}$. The internal electric field is not zero, and
generates a non-zero force within the bulk of the system. This
force is compensated by the pressure gradient $\nabla \psi$ and,
in fact, this condition determines the values of the pressure
field.


Analysis of the stability of the lamellar ground state of this
system in the absence of external fields has been analyzed by
Muratov \cite{muratov}.
Below we sketch, in notation useful for our problem, a derivation
of the stability properties of the system, ending the main result
of Eq.\ref{eq:excess}. Later, we use the results in the context of
external fields. 
The fluctuations of the system can be described by specifying the
deformations of the interfaces. We use a coordinate system ${\bf
x}=({\bf r},z)$, where the vector ${\bf r}$ runs parallel to the
lamellae and the coordinate $z$ is transversal to them. The unit
cell of the one-dimensional lattice formed by the lamellae can be
taken to be symmetric with respect to its mid-plane. We assign
$z=0$ to the midplane so that the cell extends up to $z=\pm
\lambda_{0}$ while the interfaces, labelled $\Gamma^{\pm}$, are
located at $z_{\pm}=\pm \lambda_{0}/2$. The collective
deformations of the system prescribe a transversal deformation
$\Delta {\bf u}_{n,\pm}=u_{n,\pm}\hat{\bf z}$ for each of the
interfaces $\Gamma^{\pm}_{n}$, where $n$ is an index for the cell.
The normal modes of the system have the form
$u_{n,\pm}=A_{\pm}cos(2p n\lambda_{0})\cos({\bf {\bf q\cdot r}})$,
where the wave-vector in the transversal direction ${\bf q}$ is
unrestricted while the longitudinal wavevector $p$ takes values
within the first Brillouin zone $|p|\leq \pi/(2\lambda_{0})$.
There are two normal modes for each wave vector $(p,{\bf q})$,
each with a different ratio between the deformation amplitudes
$A_{\pm}$.

\begin{figure}
\includegraphics{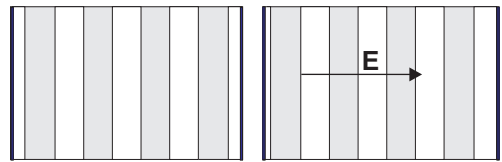}
\caption{\label{fig:ground} Ground states of a finite thickness
system. On the left, in the absence of an external filed, the
system is symmetric. The thick lines indicate the boundaries of
the system. If the unit cell width $2\lambda_{0}$ is commensurate
with the thickness, the outmost layers have equal charge and
thickness $\lambda_{0}/2$. As shown in the right frame, in the
presence of an external field the ground state is obtained by
shifting the location of the interfaces. The new charge
distribution at the boundary compensates the external field in the
bulk.}
\end{figure}


A small deformation of the interfaces has the effect of increasing
the interfacial area, and producing changes in the charge density
localized at the interface. The excess charge density interacts
with the background charge and with itself. The excess surface
tension energy density associated with a small deformation has the
form $(1/2) (\nabla_{||}u({\bf z}))^{2}$ where $\nabla_{||}$ is
the gradient along the interface. The excess charge density can be
formally written as a series expansion in powers of the
deformation amplitude. The terms of these series, $\rho^{(i)}$,
and the associated series for the the potential and field are
labelled by $i$, the order of the term in the expansion. The first
terms of the excess charge density at a given interface have the
form $\rho=\rho^{(0)}\mp 2 u({\bf r})\delta (z-z_\pm)\pm u({\bf
r})^2\delta'(z-z_\pm)$ where $\delta$ and $\delta'$ are the delta
function and its derivative. The potential and electric field can
be solved order by order in the series. The higher order terms of
the potential satisfy the Laplace equation within each homogenous
domain and have boundary conditions determined by the excess
charge density at the interfaces. The electrostatic excess energy
density is $(1/2) \phi^{(1)}\rho^{(1)}+\rho^{(2)}\phi^{(0)}$. Upon
integration of the delta functions localized at the interfaces,
the excess free energy
is the sum,
over all interfaces, of the terms
\begin{equation}
\Delta F_{n,\pm} = \int_{\Gamma} d{\bf r}
\left[\frac{1}{2}(\nabla_{\parallel} u)^{2} + \phi^{(1)} {\bf
u}\cdot\hat{\bf n}
 - u^2 {\bf E}_{i}^{ (0)} \cdot \hat {\bf n}
 \right].\label{eq:excess}
\end{equation}
We choose the orientation of the interface so that its normal
vector ${\bf n}$ points outward from the positive charge regions.
The electric field at the interfaces in the ground state is ${\bf
E}_{i}^{(0)}=2\pi \lambda_{0} \hat{\bf n}$, so that the last term
in the excess energy always has a destabilizing effect.


Carrying out the calculation sketched above, the energies of the
normal modes, as well as the relation between the deformation
amplitudes of the surfaces in the unit cell can be obtained as a
function of the wavevector $(p,{\bf q})$. An equivalent
calculation has been carried out by Muratov\cite{muratov},
including the case of asymmetric mixtures, and in the presence of
screening. The basic result is that all deformations have a
non-negative excess energy and therefore the system is stable
against small deformations. As shown below, in the presence of
constant external fields some of the modes become unstable. The
modes that become unstable are deformations of the lowest energy
modes at zero field that are located at the center, at $p=0$, and
edges $p=\pm \pi/(2\lambda_{0})$ of the first Brillouin zone. The
explicit form and energies of these modes at zero field are as
follow. For $p=0$, all unit cells have the same displacements, and
within each unit cell the normal modes are the charge symmetric
$A_{\pm}=\pm A_{s}/\sqrt{2}$, and charge antisymmetric modes
$A_{\pm}= A_{a}/\sqrt{2}$. The energy per unit volume of each of
these modes, for each transverse wavevector ${\bf q}$ with
magnitude $q$, are $(1/2)\varepsilon A^{2}$ with
\begin{equation}
\varepsilon_a(q)=2\pi\left[
\frac{\sinh(q\lambda_{0}/2)}{(q\lambda_{0}/2) \cosh(q
\lambda_{0}/2)}-1+\frac{1}{12}(q\lambda_{0})^{2}\right],
\end{equation}
\begin{equation}
\varepsilon_s(q)=2\pi\left[
\frac{\cosh(q\lambda_{0}/2)}{(q\lambda_{0}/2)
\sinh(q\lambda_{0}/2)}-1+\frac{1}{12}(q\lambda_{0})^{2}\right]
\end{equation}
These results exhibit the softness of the antisymmetric mode,
$\varepsilon_{s}\sim q^{4}$  and the divergent hardness of the
symmetric mode $\varepsilon_{s}\sim q^{-2}$, for small $q$. In the
edge normal modes with $p=\pm \pi/(2\lambda_{0})$ only one of the
unit cell surfaces is deformed. The two modes correspond to the
choice of deformed surface and clearly both modes have the same
energy density $(1/2) \varepsilon_{e} A^2$ with:
\begin{equation}
\varepsilon_e(q)=2\pi\left[
\frac{\cosh(q\lambda_{0})}{q\lambda_{0}
\sinh(q\lambda_{0})}-\frac{1}{2}+\frac{1}{12}(q
\lambda_{0})^{2}\right].
\end{equation}
The lowest energy modes in this group appear at
$q\lambda_{0}=1.6$.

Finite size effects and the action of external fields are
intimately related in this system. Before addressing the external
field, it is useful to consider the properties of the ground state
of the finite system. First, we point out that the ground state of
a slab of material of width commensurate to $2\lambda_{0}$ is
identical to the bulk solution, with lamellae parallel to the
limiting surfaces, terminating at both ends with layers of the
same charge of size $\lambda_{0}/2$. Consideration of the surface
terms generated by integration by parts during variation of the
charge density leads in principle to boundary conditions for the
system. In this case, however, these turn out to be equivalent to
the bulk equilibrium conditions, and the global minimum must be
determined from evaluation of the bulk energy restricted by the
incompressibility condition but without extra boundary
requirements.

For non-commensurate thicknesses the energy minimum is obtained
when the first two layers at both ends are still arranged
symmetrically but have thickness different from $\lambda_{0}$. If
the end layers do not have equal charge, a net electric field is
created within the system. The ground state of the finite system
in the presence of a constant perpendicular field is then simply a
lamellar structure with unbalanced end layers that create an
excess internal field that exactly compensates the external field,
as sketched in Fig. \ref{fig:ground}.

We consider now the effect of the sudden onset of an external
uniform electric field on the ground state at zero field. We first
note that the original lamellar structure is still an equilibrium
state even after the onset of an external field perpendicular to
the lamellae. Indeed, due to the incompressibility of the system,
rigid translations of the interfaces between components are not
possible and all other deformations of the interface lead, to
first order, to net zero changes in the energy of the system. In
more detail, the equilibrium of the system is achieved again by
letting the internal pressure gradient to adjust to compensate the
net electric field. This effect can always be achieved when the
external field has null curl, as is the case of an uniform field.
As we show below, however, this equilibrium state is unstable.

The external field ${\bf E}_{e\perp}$ perpendicular to the
lamellae couples to the second order term of the charge
distribution induced by deformations. The excess energy is given
by the same expression as the one obtained for the interaction of
the fluctuations with the internally generated field in Eq.
(\ref{eq:excess}). For a given surface, this energy is
\begin{equation}
\Delta F=-\int_{\Gamma}u^{2}{\bf E}_{\perp e}\cdot \hat{\bf n}.
\end{equation}
This interaction has a destabilizing effect only in the interfaces
whose normal is parallel to the external field. When the
deformations are expressed in terms of zero-field normal modes,
the field couples to a mixture of the $p=0$ symmetric and
antisymmetric modes. The excess energy density is $\Delta f =
E_{e{\perp}}(A_{-}({\bf q})^{2}-A_{+}({\bf
q})^{2})/(4\lambda_{0})=- A_{s}A_{a}E_{e\perp}/\lambda_{0}$. The
new $p=0$ normal modes in the presence of the field are obtained
from diagonalization of the new energy quadratic form. These new
modes are smooth modifications of the zero-field modes, and will
be labelled by the corresponding zero-field labels $a$ or $s$.
While it is clear that non-zero average excess energies arise for
$p=0$ modes, the coupling to the square of the amplitude also
gives non-vanishing contributions for the modes at the edge of the
Brillouin zone at $p=\pm \pi/(2\lambda_{0})$. The external field
shifts the edge modes by
$\Delta f=\pm A^{2}E_{e\perp}/(2\lambda_{0})$.  

\begin{figure}
\includegraphics{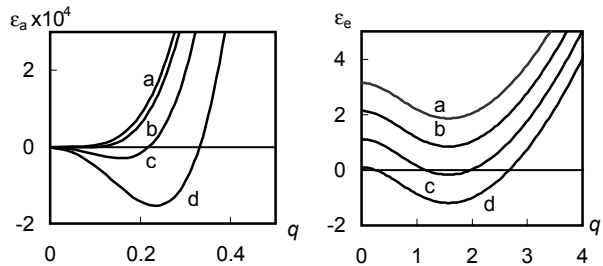}
\caption{\label{fig:coef} Plots of the coefficients of the energy
quadratic form as a function of the transversal wave vector
magnitude $q$. Results for the $p=0$ charge antisymmetric modes
are shown on the left, and Brillouin zone edge modes with
$p=\pi/(2\lambda_{0})$ on the right. The labels correspond in both
cases to the magnitude of the external transversal electric field
(a) $E_\perp=0$, (b) $E_\perp=0.5$,(c) $E_\perp=1.0$, and (d)
$E_\perp=1.5$. Unstable modes are those with $\varepsilon<0$.}
\end{figure}

There are two sets of unstable modes in the system in the presence
of an external perpendicular field. The modes obtained from the
deformation of the antisymmetric charge mode, with $p=0$, become
unstable for any non-zero value of the external field. These
unstable modes appear within a wavevector disk, centered at $q=0$,
of radius $q_{r} \sim E_{e}$. The most unstable wavevector has
$q_{max} \sim E_{e}$ and the energy coefficient of this mode
scales as $\varepsilon \sim -E^{8}_{e}$. Therefore at very low
field strengths the instability is very weak while, at stronger
fields, the dynamics are dominated by the exponential growth of
finite wavelength modes. A band of edge modes around the softest
edge mode, located at $p=\pi/(2\lambda_{0})$, $q_{se}=1.6$, also
becomes unstable at the finite field value $E_{ec}=0.9$. Above
this threshold value the band of unstable modes has width $\Delta
q \sim (E-E_{ec})^{1/2}$. The most unstable mode is always
$q_{se}$, with energy coefficient
$\varepsilon_{e}(q_{ec})-E/\lambda_{0}$. Plots of the energy
coefficients for these modes appear in Fig. \ref{fig:coef}.

\begin{figure}
\includegraphics{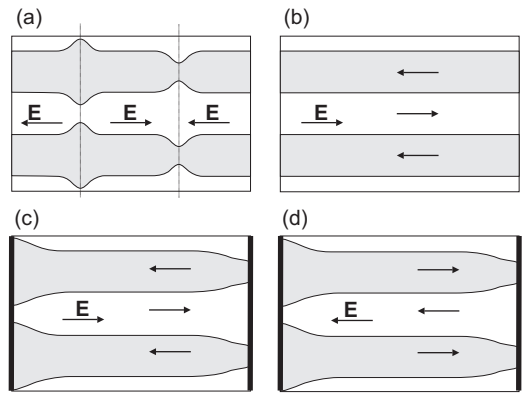}
\caption{\label{fig:dyn} Effect of uniform fields parallel to the
lamellae. (a) Periodically alternating
uniform fields cause a concentration of charge at the ends of the
uniform region. (b) In an infinite system, steady currents are
possible; the arrows indicate flow direction. (c) In finite
systems the field induces smooth flows. (d) Reversal of the field
simply reverses the flow but does not
destabilize it.}
\end{figure}

We consider now the effect of a field ${\bf E}_{\parallel}$
parallel to the lamellae. The ultimate end effect of this field
will be to orient the lamellae in a direction {\it perpendicular}
to the field. There are, however, dynamical steady states with
orientations parallel to the field. To show this, we decompose the
external potential into its Fourier components $\hat\phi_{\bf q}$.
These components couple linearly to the symmetric deformation
modes of same wavevector and their excess energy density in this
case is $\Delta f=(1/2)\varepsilon_{s}({\bf q})A_{s}( {\bf
q})^{2}+\sqrt{2}\hat\phi_{\bf q}A_{s}({\bf q})/\lambda_{0}$.
Therefore, starting from a parallel orientation, the system flows
to a metastable state with symmetric deformations of amplitude
$A_s({\bf q})=\sqrt{2}\hat\phi_{\bf q} /(\lambda
_{0}\varepsilon_{s}({\bf q}))$. In a system where the external
field is locally constant but periodic, with periodicity $L$ in
the direction of the lamellae (see Fig. \ref{fig:dyn}), the
potential has non-vanishing Fourier components of the form
$\hat\phi_{\bf q} \sim {E_{\parallel}}q^{-2}$. We pointed out
above a similar behavior for the self-energy of the long
wavelength symmetric fluctuations $\varepsilon_{s}\sim q^{-2}$.
The deformation amplitudes are then near uniform for small
wavectors, and very small for large ones. The net result for large
periodicities $L$ is an accumulation of charge at the edges of
each homogeneous region as in Fig. \ref{fig:dyn}a. These
deformations require net charge exchanges between the boundaries
of the homogenous field regions. In the limit of infinite size,
this charge redistribution emerges as steady exchange currents as
sketched in Fig. \ref{fig:dyn}b.



The ground states corresponding to different external fields
cannot be connected by charge redistributions that preserve the
integrity of the layers since there are no smooth deformations
that change the charge of the end layers. Therefore, the flow
between ground states always occurs by means of the destruction of
the original structure by the growth of instabilities.
 The flow between the ground states for different
external fields requires net transport of charge which, as noted
above, occurs naturally in lamellar orientations parallel to the
external field. After the destruction of the layers parallel to
the boundary, charge transfer might occur through the formation of
layers perpendicular to the boundary but parallel to the field,
until the charge unbalance necessary to shield the bulk from the
external field is achieved. Afterwards, the intermediate layers
may become unstable and flow into the final stable conformation
with layers parallel to the boundary.


While the presence of instabilities precludes the smooth flow
between ground states, i.e. adiabatic control of the system by
tuning of external fields, there are interesting possibilities for
the creation of structures through slowly varying uniform fields.
In particular, we note that smooth deformations of lamellae under
parallel fields carry out a similar process as a growing unstable
mode in a perpendicular field; namely, both transfer charge along
the field direction. That is, from an energetic point of view, the
preferred intermediate states between the ground states of a slab
with perpendicular fields are those that connect same-charge
regions across the slab. When driven by an oscillating uniform
field, the system may oscillate along a set of states of easy
charge transfer, as sketched in Fig. \ref{fig:dyn}(c-d), within a
suitable range of frequencies determined by the specific dynamics
of the system. In other words, oscillating fields may dynamically
stabilize lamellar orientations parallel to them.

In conclusion, we have shown how to use the static stability
analysis of the system as a guide to describe possible dynamic
scenarios for the flow of the system towards equilibrium. The key
result is that lamellar structures are unstable equilibrium states
in the presence of external perpendicular electric fields, while
fields parallel to the lamellae give rise to natural charge
transferring currents.

\bibliography{revised_instabilities}
\end{document}